\documentclass{elsarticle}
\usepackage{amssymb}
\usepackage{amsmath}
\usepackage{color}
\usepackage{multicol}
\usepackage{longtable}
\usepackage{booktabs}

\usepackage{graphicx}
\usepackage{hyperref}
\usepackage{xcolor}
\hypersetup{colorlinks=true, linkcolor=blue, citecolor=blue}
\usepackage[toc,page]{appendix}

\usepackage{placeins}
\usepackage{multirow}
\usepackage{soul}%\st to remove, \ul to underline

\newcommand{\+}{\hspace{-1mm}+\hspace{-1mm}}

\begin{document}

%%%%%%%%%%%%%%%%%%%%%%%%%%%%%%%%%%%%%%%%%%%%%%%%%%%%%%%%%%%%%%%%%%%%%%%%%%%%%%%%%%
%%%%%%%%%%%%%%%%%%%%%%%%%%%%%%%%%%%%%%%%%%%%%%%%%%%%%%%%%%%%%%%%%%%%%%%%%%%%%%%%%%
\begin{frontmatter}

\title{Impact of unitarization on the $J/\psi$-light meson cross section}

\author[UFBA]{L. M. Abreu\corref{cor1}}
\ead{luciano.abreu@ufba.br}
\cortext[cor1]{Corresponding author}
\author[CBPF]{E. Cavalcanti}
\ead{erich@cbpf.br}
\author[CBPF]{A. P. C. Malbouisson}
\ead{adolfo@cbpf.br}

\address[UFBA]{Instituto de F\'{\i}sica, Universidade Federal da
Bahia, 40170-115, Salvador, BA, Brazil}

\address[CBPF]{Centro Brasileiro de Pesquisas F\'isicas/MCTI, 22290-180, Rio de Janeiro, RJ, Brazil}
%

%\author{L.M. Abreu}
%
%\email{luciano.abreu@ufba.br}
%
%\affiliation{Instituto de F\'isica, Universidade Federal da Bahia, 40210-340, Salvador, BA, Brazil}
%
%
%
%\author{E. Cavalcanti}
%
%\email{erich@cbpf.br}
%
%\affiliation{Centro Brasileiro de Pesquisas F\'isicas/MCTI, 22290-180, Rio de Janeiro, RJ, Brazil}
%
%
%
%\author{A.P.C. Malbouisson}
%
%\email{adolfo@cbpf.br}
%
%\affiliation{Centro Brasileiro de Pesquisas F\'isicas/MCTI,	22290-180, Rio de Janeiro, RJ, Brazil}
%
%
%

%%%%%%%%%%%%%%%%%%%%%%%%%%%%%%%%%%%%%%%%%%%%%%%%%%%%%%%%%%%%%%%%%%%%%%%%%%%%%%%%%%
%%%%%%%%%%%%%%%%%%%%%%%%%%%%%%%%%%%%%%%%%%%%%%%%%%%%%%%%%%%%%%%%%%%%%%%%%%%%%%%%%%
\begin{abstract}

Hidden charm mesons continue playing an essential role as relevant probes to understand the evolution of partonic matter. 
It is expected that the charmonia that survived the quark-gluon plasma phase suffer collisions with other particles composing the hadronic matter. In this work, we intend to contribute on this subject by presenting an updated study about the interactions of $J/\psi$ with surrounding hadronic medium. The meson-meson interactions are described with a $SU(4)$ effective Lagrangian, and within the framework of unitarized coupled channel amplitudes projected onto $s$-wave. The symmetry is explicitly broken to  $SU(3)$ by suppression of the interactions driven by charmed mesons. We calculate the cross sections for $J/\psi $ scattering by light pseudoscalar mesons ($\pi, K, \eta$) and vector mesons ($\rho, K^\ast, \omega$), as well as their inverse processes. Keeping the validity of this present approach in the low CM energy range, the most relevant channels are evaluated and a comparison of the findings with existing literature is performed.

\end{abstract}

\end{frontmatter}
%%%%%%%%%%%%%%%%%%%%%%%%%%%%%%%%%%%%%%%%%%%%%%%%%%%%%%%%%%%%%%%%%%%%%%%%%%%%%%%%%%
%%%%%%%%%%%%%%%%%%%%%%%%%%%%%%%%%%%%%%%%%%%%%%%%%%%%%%%%%%%%%%%%%%%%%%%%%%%%%%%%%%

%\maketitle

%%%%%%%%%%%%%%%%%%%%%%%%%%%%%%%%%%%%%%%%%%%%%%%%%%%%%%%%%%%%%%%%%%%%%%%%%%%%%%%%%%
%%%%%%%%%%%%%%%%%%%%%%%%%%%%%%%%%%%%%%%%%%%%%%%%%%%%%%%%%%%%%%%%%%%%%%%%%%%%%%%%%%
%%%%%%%%%%%%%%%%%%%%%%%%%%%%%%%%%%%%%%%%%%%%%%%%%%%%%%%%%%%%%%%%%%%%%%%%%%%%%%%%%%
%%%%%%%%%%%%%%%%%%%%%%%%%%%%%%%%%%%%%%%%%%%%%%%%%%%%%%%%%%%%%%%%%%%%%%%%%%%%%%%%%%
\section{Introduction} 
\label{Introduction} 
%%%%%%%%%%%%%%%%%%%%%%%%%%%%%%%%%%%%%%%%%%%%%%%%%%%%%%%%%%%%%%%%%%%%%%%%%%%%%%%%%%
%%%%%%%%%%%%%%%%%%%%%%%%%%%%%%%%%%%%%%%%%%%%%%%%%%%%%%%%%%%%%%%%%%%%%%%%%%%%%%%%%%
%%%%%%%%%%%%%%%%%%%%%%%%%%%%%%%%%%%%%%%%%%%%%%%%%%%%%%%%%%%%%%%%%%%%%%%%%%%%%%%%%%
%%%%%%%%%%%%%%%%%%%%%%%%%%%%%%%%%%%%%%%%%%%%%%%%%%%%%%%%%%%%%%%%%%%%%%%%%%%%%%%%%%

Recent heavy-ion-collision experiments generated a prosperous era in particle and nuclear physics. Measurements that seemed hard to be performed two or three decades ago can now be done with unprecedent precision.
Among them, those related to heavy-flavored hadrons have been proved to play an essential role.  
These states are of particular interest since they carry heavy quarks produced by hard gluons in the initial stages of collisions. Noticing that the hadronic medium is not hot enough to excite heavy-quark pairs, heavy hadrons are relevant probes to understand the evolution of partonic matter, in contrast to light hadrons, which can be yielded in the thermal medium at later stages.

In this scenario, the $J/\psi$ reveals itself as a relevant probe of properties of quark-gluon plasma (QGP) phase produced in the collision. It relies on the suggestion done about three decades ago that this phase  would screen the $c-\bar{c}$ interaction, leading to the drop of $J/\psi$ multiplicity~\cite{Matsui:1986dk,RAPP2010209,Braun-Munzinger}.
Indeed, several Collaborations have observed experimental evidences of $J/\psi$ suppression~\cite{Gonin:1996wn,Abreu:1997jh,Alessandro:2004ap,Arnaldi:2006ee,Arnaldi:2007zz,Adare:2006nq}. However, at the highest energies reached today at the LHC, data on $J/\psi$ production confirm that the QGP dynamics is richer and more complex. 
At low transverse momentum ($p_T$) range, the $J/\psi$ drop is significantly smaller at LHC energy than at RHIC energy, which might be interpreted from regeneration mechanism due to larger total charm cross section at LHC; but at high $p_T$ the dissociation increases as collision energy grows, indicating that the  $J/\psi$ yield is less sensitive to recombination and other effects~\cite{Abelev:2013ila,Adam:2016rdg,Zha:2017xsm}. 
 
On the other hand, alternative mechanisms have also been proposed to explain the drop of charmonium multiplicity, such as its absorption by comoving hadrons. 
It is worthy mentioning that between the chemical freeze-out (where the hadronization has already ended and there is a hadron gas) and the kinetical freeze-out (in which the interactions are expected to cease and the remaining particles go to the detectors), the charmonia that have survived the QGP phase are expected to collide with other particles composing the hadronic matter. Therefore, inelastic interactions of $J/\psi$ with surrounding hadronic medium formed after QGP cooling and hadronization might have (at least partially) significance on the charmonium abundance analysis. 
 
In this sense, a large amount of effort has been dedicated to estimate the charmonia interactions with light hadrons (mainly involving $\pi$ and $\rho$ mesons) using different approaches~\cite{Wong:1999zb,Wong:2001td,PhysRevC.58.2994,PhysRevC.61.031902,Braun-Munzinger2000,PhysRevC.62.034903,PhysRevC.63.065201,PhysRevC.63.034901,PhysRevC.68.014903,Oh:2002vg,PhysRevC.68.035208,Maiani:2004py,Maiani:2004qj,DURAES200397,PhysRevC.70.055203,PhysRevC.72.024902,PhysRevD.72.034002,Capella2008,CASSING20011,doi:10.1142/S0218301308010507,PhysRevC.85.064904,MITRA201675,Liu2016,PhysRevC.96.045201,PhysRevC.97.044902}. 
%In some of them  medium and finite-temperature effects are explicitly included~\cite{PhysRevC.62.034903,MITRA201675,Liu2016,PhysRevC.96.045201,PhysRevC.97.044902}.
Most of these analyses explore the $J/\psi - \pi$ reactions with reasonable results, and can be classified in the following sort: interactions based on effective hadron Lagrangians~\cite{PhysRevC.58.2994,PhysRevC.61.031902,Braun-Munzinger2000,PhysRevC.62.034903,PhysRevC.63.065201,PhysRevC.63.034901,Oh:2002vg,PhysRevC.70.055203,PhysRevC.72.024902,MITRA201675,PhysRevC.96.045201,PhysRevC.97.044902} and constituent quark-model framework~\cite{Wong:1999zb,Wong:2001td,Braun-Munzinger2000,PhysRevC.68.014903,Maiani:2004py,Maiani:2004qj,PhysRevC.85.064904,Liu2016}.

Concerning those works involving $J/\psi$ absorption by light hadrons (and their inverse reactions) derived from chiral Lagrangians, we believe that there is still enough room for other contributions on this issue. First, due to the fact that the charmonium-hadron cross sections are dependent of the effective couplings that control the reactions considered~\cite{PhysRevC.58.2994,PhysRevC.61.031902,Braun-Munzinger2000,PhysRevC.62.034903,PhysRevC.63.065201,PhysRevC.63.034901,Oh:2002vg,PhysRevC.70.055203,PhysRevC.72.024902,MITRA201675,PhysRevC.96.045201,PhysRevC.97.044902}. Secondly, the majority of these mentioned calculations make use of form factors with different functional forms and cutoff values which could not be justified a priori. It should be also mentioned that appropriate choice for the form factors is essential to obtain reliable predictions, since the range of heavy meson exchange is much smaller than the sizes of the initial hadrons~\cite{Oh:2002vg}.
Third, the older calculations are deficient of the methods that have been developed subsequently, as well as lack the novel data of heavy-ion-collision experiments at RHIC and LHC, which requires a new round of updated predictions. 

Thus, in the present work we will contribute on calculations about the interactions of $J/\psi$ with surrounding hadronic medium compared to previous studies in the following way. We consider the medium composed of the lightest pseudoscalar mesons ($\pi, K, \eta$) and the lightest vector mesons ($\rho, K^\ast, \omega$), and calculate the cross sections for $J/\psi X $ scattering and their inverse processes (in which $X$ stands for light pseudoscalar and vector mesons), within the framework of unitarized coupled channel amplitudes projected onto $s$-wave~\cite{PhysRevC.96.045201,Roca:2005nm,Gamermann:2007fi,Abreu2011,Abreu2013a}. We analyze the magnitude of unitarized cross sections of the different channels, and perform a comparison of our results with other reported ones.

This work is structured as follows. In Section II we will give an overview of the effective $SU(4)$ model and calculate the unitarized coupled channel amplitudes.  Results will be presented in Section III. We summarize the results and conclusions in Section IV. Some relevant tables  are given in Appendix A.

%%%%%%%%%%%%%%%%%%%%%%%%%%%%%%%%%%%%%%%%%%%%%%%%%%%%%%%%%%%%%%%%%%%%%%%%%%%%%%%%%%
%%%%%%%%%%%%%%%%%%%%%%%%%%%%%%%%%%%%%%%%%%%%%%%%%%%%%%%%%%%%%%%%%%%%%%%%%%%%%%%%%%
%%%%%%%%%%%%%%%%%%%%%%%%%%%%%%%%%%%%%%%%%%%%%%%%%%%%%%%%%%%%%%%%%%%%%%%%%%%%%%%%%%
%%%%%%%%%%%%%%%%%%%%%%%%%%%%%%%%%%%%%%%%%%%%%%%%%%%%%%%%%%%%%%%%%%%%%%%%%%%%%%%%%%
\section{Formalism}
\label{formalism} 
%%%%%%%%%%%%%%%%%%%%%%%%%%%%%%%%%%%%%%%%%%%%%%%%%%%%%%%%%%%%%%%%%%%%%%%%%%%%%%%%%%
%%%%%%%%%%%%%%%%%%%%%%%%%%%%%%%%%%%%%%%%%%%%%%%%%%%%%%%%%%%%%%%%%%%%%%%%%%%%%%%%%%
%%%%%%%%%%%%%%%%%%%%%%%%%%%%%%%%%%%%%%%%%%%%%%%%%%%%%%%%%%%%%%%%%%%%%%%%%%%%%%%%%%
%%%%%%%%%%%%%%%%%%%%%%%%%%%%%%%%%%%%%%%%%%%%%%%%%%%%%%%%%%%%%%%%%%%%%%%%%%%%%%%%%%

The main purpose here is the discussion of $J/\psi$ interaction with the hadronic medium. We intend to calculate and analyze the cross sections
for the $J/ \psi - X$ interactions, where $X$ denotes a pseudoscalar or vector meson. On that subject, we work within the framework of effective field theories whose hadrons are the relevant degrees of freedom. The effective Lagrangian used in the present study is based on $SU(4)$ lowest order Chiral Perturbation Theory~\cite{PhysRevC.96.045201,Roca:2005nm,Gamermann:2007fi},
\begin{equation}
\mathcal{L}_{\text{int}} = -\frac{1}{4f^2} \text{Tr}\left(J^\mu \mathcal{J}_\mu \right) -\frac{1}{4f^2} \text{Tr}\left(\mathcal{J}^\mu \mathcal{J}_\mu \right),
\label{Lagr}
\end{equation}
where $Tr (...)$ denotes the trace over flavor indices, $J^\mu = [P,\partial^\mu P]$ and $\mathcal{J}^\mu = [V^\nu, \partial^\mu V_\nu]$ are the pseudoscalar and vector currents, respectively, with $P$ and $V$ being $4 \times 4$ matrices carrying 15-plets of pseudoscalar and vector fields as show below in an unmixed representation,
%\footnote{Other Refs.\cite{Abreu:2017cof} may choose to work with a mixed representation.}
%\begin{widetext}
\begin{eqnarray}
&  &  P = 
 \sum_{i=1}^{15} \frac{\varphi_i}{\sqrt{2}} \lambda_i=  \nonumber \\
& & \begin{pmatrix}
\frac{\pi^0}{\sqrt{2}}\+\frac{\eta}{\sqrt{6}} \+ \frac{\eta_c}{\sqrt{12}} & \pi^+ & K^+ & \bar D^0\\
\pi^- & -\frac{\pi^0}{\sqrt{2}}\+\frac{\eta}{\sqrt{6}}\+\frac{\eta_c}{\sqrt{12}} & K^0 & D^-\\
K^- & \bar K^0 & -2\frac{\eta}{\sqrt{6}}\+\frac{\eta_c}{\sqrt{12}} & D_s^-\\
D^0 & D^+ & D_s^+ & -\frac{\sqrt{3}}{2} \eta_c
\end{pmatrix}; \,\,\,
\nonumber \\
& & V_\mu  =   
\sum_{i=1}^{15} \frac{v_{\nu i}}{\sqrt{2}} \lambda_i= \nonumber \\
& & \begin{pmatrix}
\frac{\rho^0}{\sqrt{2}}\+\frac{\omega}{\sqrt{6}}\+\frac{J/\psi}{\sqrt{12}} & \rho^+ & K^{*+} & \bar D^{*0}\\
\rho^- & -\frac{\rho^0}{\sqrt{2}}\+\frac{\omega}{\sqrt{6}}\+\frac{J/\psi}{\sqrt{12}} & K^{*0} & D^{*-}\\
K^{*-} & \bar K^{*0} & -2\frac{\omega}{\sqrt{6}}\+\frac{J/\psi}{\sqrt{12}} & D_s^{*-}\\
D^{*0} & D^{*+} & D_s^{*+} & -\frac{\sqrt{3}}{2} J/\psi
\end{pmatrix}_\mu; \nonumber \\
\label{PV}
\end{eqnarray}
%\end{widetext}
$\lambda_a$ being the Gell-Mann matrices for $SU(4)$.
The parameter $f$ is the meson decay constant, which is the pion decay constant in the usual $SU(3)$ symmetry. But here $f^2$ which will appear in the amplitudes must be replaced by $\sqrt{f}$ for each meson leg in the corresponding vertex, with $\sqrt{f_{\pi}}$ for light mesons and $\sqrt{f_D}$ for heavy ones.

The couplings given by the effective Lagrangian in Eq.~(\ref{Lagr}) allows us to obtain the scattering amplitudes for the following $ J/ \psi X$ absorption processes: 
\begin{eqnarray}
(1) \;\; J/ \psi (p_1) P (p_2) & \rightarrow &  V (p_3) P (p_4) , \nonumber \\
(2) \;\; J/ \psi (p_1) V (p_2) & \rightarrow &  P (p_3) P (p_4) , \nonumber \\
(3) \;\;J/ \psi (p_1) V (p_2) & \rightarrow &  V (p_3) V (p_4) ,
\label{proc1}
\end{eqnarray}
where $P$ and $V$ in the initial and final states stand for pseudoscalar and vector mesons, and $p_j$  denotes the momentum of particle $j$, with particles 1 and 2 standing for initial state mesons, and particles 3 and 4 for final state mesons. 

Thus, the invariant amplitudes engendered by effective Lagrangian in Eq.~(\ref{Lagr}) for processes of type $ V P\rightarrow V P $ in Eq.~(\ref{proc1}) are given by
\begin{eqnarray}
\mathcal{M} _{1; i j}  (s, t, u )= \frac{\xi _{i j}  }{2f^2} (s-u)  \varepsilon _1 \cdot\varepsilon _3 ^{\ast} ,
\label{Eq:CasoVPVP}
\end{eqnarray}
for processes $V V \rightarrow P P$ they are
\begin{eqnarray}
\mathcal{M} _{2; i j}  (s, t, u )= \frac{ \chi _{i j}  }{2f^2} (t-u)  \varepsilon _1 \cdot\varepsilon _2 ,
\label{Eq:CasoVVPP}
\end{eqnarray}
and finally for processes $V V \rightarrow V V$,
\begin{eqnarray}
\mathcal{M} _{3; i j}   (s, t, u ) & = & 
\frac{\zeta _{ i j } ^{(s)} }{f^2} (t-u) \varepsilon _1 \cdot\varepsilon _2 \varepsilon ^{\ast} _3 \cdot\varepsilon _4 ^{\ast} \nonumber \\
& & + \frac{ \zeta _{ i j } ^{(t)} }{f^2} (s-u)  \varepsilon_1\cdot\varepsilon _3 ^{\ast}  \varepsilon _2 \cdot\varepsilon _4 ^{\ast} \nonumber \\
& & + \frac{\zeta _{ i j } ^{(u)} }{f^2} (s-t)  \varepsilon _1 \cdot\varepsilon _4 ^{\ast}  \varepsilon _2 \cdot\varepsilon _3 ^{\ast}, 
\label{Eq:CasoVVVV}
\end{eqnarray}
where the labels $i$ and $j$ refer to the initial
and final channels; $s, t$ and $u$ to the Mandelstam variables; $\varepsilon _a$ to the polarization vector related to the respective vector particle $a$.
The coefficients $\xi _{i j}, \chi _{i j } $ and $\zeta _{i j}$ will depend on the initial and final channels of each process, and are given in~\ref{Appendix} in an isospin basis.

The processes above are assumed to have conservation of the quantum numbers for the incoming and outcoming meson pairs; they are $I^G(J^{PC})$, charm ($C$) and strangeness ($S$). Therefore, relating to $s$-wave reactions, we deal with the channels involving pairs of vector mesons in Eq.~(\ref{Eq:CasoVVVV}) by making use of spin-projectors that distinguish the allowed values of spin~\cite{Roca:2005nm,PhysRevD.78.114018}. Explicitly, suppose a given generic amplitude,
\begin{eqnarray}
\mathcal{A} & = & \alpha \,\varepsilon _1 \cdot\varepsilon _2 \varepsilon ^{\ast} _3 \cdot\varepsilon _4 ^{\ast} 
+ \beta \, \varepsilon_1\cdot\varepsilon _3 ^{\ast}  \varepsilon _2 \cdot\varepsilon _4 ^{\ast} \nonumber \\
& & + \gamma \, \varepsilon _1 \cdot\varepsilon _4 ^{\ast}  \varepsilon _2 \cdot\varepsilon _3 ^{\ast}.
\label{ampA1}
\end{eqnarray}
We can decompose the polarization vectors of each incoming/outgoing pair of vector mesons into the following representations: scalar ($S=0$), antisymmetric tensor ($S=1$) and symmetric tensor ($S=2$), namely
\begin{eqnarray}
\varepsilon_a ^{i} \varepsilon_b ^{j} = \mathcal{P}_{a b } ^{(S=0)i j } +\mathcal{P}_{a b } ^{(S=1) i j } + \mathcal{P}_{a b } ^{(S=2) i j }
\label{ampA2}
\end{eqnarray}
where
\begin{eqnarray}
\mathcal{P}_{a b } ^{(S=0)i j }  &  = & \frac{\delta^{ij}}{3} \varepsilon_a ^{k} \varepsilon_b^{k}, 
\nonumber \\
\mathcal{P}_{a b } ^{(S=1) i j }  &  = & \frac{1}{2}\left(\varepsilon_a ^{i} \varepsilon_b ^{j} - \varepsilon_a ^{j} \varepsilon_b ^{i}\right) \nonumber \\
\mathcal{P}_{a b } ^{(S=2) i j }  &  = & \frac{1}{2}\left(\varepsilon_a ^{i} \varepsilon_b ^{j} + \varepsilon_a ^{j} \varepsilon_b ^{i}\right)-\frac{\delta^{ij}}{3} \varepsilon_a ^{k} \varepsilon_b^{k}. 
\label{ampA3}
\end{eqnarray}
Then, using this decomposition in Eq.~(\ref{ampA1}), the generic amplitude can be written as
\begin{equation}
\mathcal{A} = (3\alpha+\beta+\gamma) \mathcal{A}^{(S=0)} + (\beta-\gamma) \mathcal{A}^{(S=1)} + (\beta+\gamma) \mathcal{A}^{(S=2)}.
\label{ampA4}
\end{equation}
where
\begin{eqnarray}
\mathcal{A}^{(S=0)} & \equiv & \mathcal{P}_{a b} ^{(S=0) i i }\mathcal{P}_{c d} ^{(S=0) j j }, \nonumber \\
\mathcal{A}^{(S=1)}  & \equiv & \mathcal{P}_{a b } ^{(S=1)i j }\mathcal{P}_{c d }^{(S=1) i j } ,\nonumber \\
\mathcal{A}^{(S=2)}  & \equiv & \mathcal{P}_{a b} ^{(S=2) i j } \mathcal{P}_{c d} ^{(S=2) i j } .
\label{ampA5}
\end{eqnarray}
Hence, the coefficients in the amplitude depends on the total angular momentum. We also remark that for $VV \rightarrow PP$ reactions in Eq.~(\ref{Eq:CasoVVPP}), the only relevant contribution comes from $\mathcal{P}_{a b} ^{(S=0)}$.

In order to have the correct behavior of the amplitudes at high energies, we need to implement a control procedure of the energy-dependence of cross sections. As mentioned before, most calculations found in literature for some reactions of our interest make use of form factors with different functional forms and cutoff values which could not be justified a priori~\cite{PhysRevC.58.2994,PhysRevC.61.031902,Braun-Munzinger2000,PhysRevC.62.034903,PhysRevC.63.065201,PhysRevC.63.034901,Oh:2002vg,PhysRevC.70.055203,PhysRevC.72.024902,MITRA201675,PhysRevC.96.045201,PhysRevC.97.044902}.

We adopt another scheme in the present approach: we work within the framework of unitarized coupled channel amplitudes. It ensures the validity of the optical theorem and enhances the range of applicability of the effective model controlling the behavior of the amplitudes at large energies, and has properly described hadronic resonances and meson-meson scattering~\cite{PhysRevC.96.045201,Roca:2005nm,Gamermann:2007fi,Abreu2011,Abreu2013a,Weinstein:1990gu,Janssen:1994wn,Oller:1997ti,Oller:1998hw}.

\begin{figure}[h]
	\centering
	\includegraphics[width=1.0\linewidth]{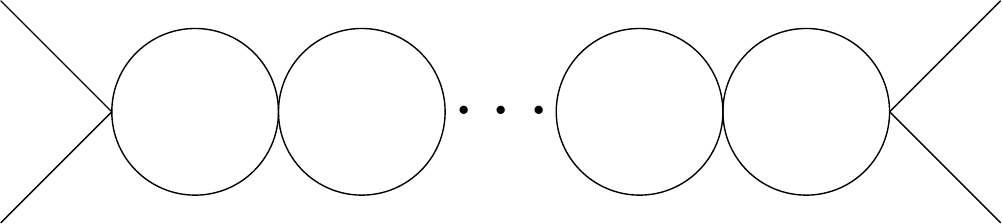}
	\caption{Feynman Diagrams representing the Bethe-Salpeter equation for the  scattering amplitudes. Each loop denotes a two-meson loop integral $G$.}
	\label{fig:subprocess}
\end{figure}

The matrix representing unitarized coupled channel transitions can be derived by a  Bethe-Salpeter equation whose kernel is the $s$-wave projection of a given amplitude by Eqs.(\ref{Eq:CasoVPVP}), (\ref{Eq:CasoVVPP}) or (\ref{Eq:CasoVVVV}), and can be diagrammatically viewed as the sum over processes showed in Fig.~\ref{fig:subprocess}. In this way, the unitarized amplitude reads~\cite{PhysRevC.96.045201,Roca:2005nm,Gamermann:2007fi,Abreu2011,Abreu2013a,Weinstein:1990gu,Janssen:1994wn,Oller:1997ti,Oller:1998hw},
\begin{equation}
\mathcal{T} (s)= \frac{V (s)}{1 + V(s) G(s)},
\label{Tmatrix}
\end{equation} 
where $V(s)$ is the s-wave projected scattering amplitude, 
\begin{equation}
V_{r; i j } (s) = \frac{1}{2} \int_{-1}^{1} d(\cos \theta) \mathcal{M} _{r; i j} \left( s, t(s,\cos \theta), u(s,\cos \theta)\right),
\label{Vamp}
\end{equation} 
with $r=1,2,3$, and $G (s)$ stands for the two-meson loop integral. In the case of two pseudoscalars mesons (PP), $G_{PP}(s)$ is given by
\begin{eqnarray}
G_{PP}(s) &= & i \int \frac{d^4 q}{(2\pi)^4} \frac{1}{\left(q^2-m_1^2+i\epsilon\right)\left[(P-q)^2-m_2^2+i \epsilon \right]}. \nonumber \\
\label{G1}
\end{eqnarray}
$ P^2 = s $ and $m_1$ and $ m_2$ are pseudoscalar mesons masses. Employing dimensional regularization, this integral is rewritten as 
\begin{eqnarray}
G_{PP}(s) & = & \frac{1}{16 \pi^2} \left\{ a(\mu) + \ln \frac{m_1^2}{\mu^2} + \frac{m_2 ^2-m_1^2 + s}{2s} \ln \frac{m_2^2}{m_1^2} \right.\nonumber\\ 
& &  + \frac{p}{\sqrt{s}} \left[
\ln(s-(m_1^2-m_2^2)+2 p\sqrt{s})\right. \nonumber \\ 
& & + \ln(s+(m_1^2-m_2^2)+2 p\sqrt{s}) \nonumber  \\ 
& & - \ln(s-(m_1^2-m_2^2)-2 p\sqrt{s}) \nonumber \\  
& & \left. \left. 
-\ln(s+(m_1^2-m_2^2)-2 p\sqrt{s}) - 2 \pi i \right] 
\right\} \ ,
\label{G2}
\end{eqnarray}
where $\mu$ is the regularization energy scale, $a(\mu)$ is a subtraction
constant which absorbs the scale dependence of the integral, and $p$ is the three-momentum in the center of mass frame of the two mesons in channel $PP$,
\begin{eqnarray}
p = \frac{1}{2\sqrt{s}} \sqrt{\left[ s - (m_{1} + m_2 )^2 \right] \left[s - (m_{1} -m_2)^2 \right]} .
\end{eqnarray}
When the two-meson loop integral involves a pseudoscalar and a vector meson (PV) and two vector mesons (VV),  we perform standard approximation as in previous studies~\cite{Roca:2005nm,Gamermann:2007fi}, resulting in the expressions
\begin{eqnarray}
G_{VP} (s) &= & \left(1 + \frac{p^2}{3 M_1^2}\right) G_{PP}(s), \nonumber \\
G_{VV} (s) &= & \left(1 + \frac{p^2}{3 M_1^2}\right) \left(1 + \frac{p^2}{3 M_2^2}\right) G_{PP}(s).
\label{G3}
\end{eqnarray}
where $M_1$ and $M_2$ represent the masses of vector mesons in the loop. Notice that   the masses in $G_{PP}(s)$ that appear in Eq.~(\ref{G3}) must be replaced by the masses of the mesons in the loop according to each case.

Once the unitarized transition amplitudes are obtained, we can determine the isospin-spin-averaged cross section for the processes in Eq. (\ref{proc1}), which in the center of mass (CM) frame  is  defined as
\begin{equation}
\sigma(s) =  \frac{ \chi }{32 \pi s}\overline{\sum_{\text{Isospin}}}\left|\frac{p_f}{p_i}\right||\mathcal{T} (s)|^2 .
\label{eq:CrossSection}
\end{equation}

\noindent where $p_f$ and $p_i$ are, respectively the momentum of the outcoming and incoming particles in the CM frame; $\chi$ is a constant whose value depends on the total angular momentum of the channel considered:
\begin{eqnarray*}
\chi = 2 &\quad& (PP \rightarrow PP, VP \rightarrow VP)\\
\chi = 6 &\quad& (PP \rightarrow VV)\\
\chi = 2/3 &\quad& (VV \rightarrow PP)\\
\chi = 2/9 &\quad& (VV \rightarrow VV; S=0)\\
\chi = 2/3 &\quad& (VV \rightarrow VV; S=1)\\
\chi = 10/9 &\quad& (VV \rightarrow VV; S=2).
\end{eqnarray*}

Next, we use the formalism developed above to compute the cross sections of reactions involving charmonium.

%%%%%%%%%%%%%%%%%%%%%%%%%%%%%%%%%%%%%%%%%%%%%%%%%%%%%%%%%%%%%%%%%%%%%%%%%%%%%%%%%%
%%%%%%%%%%%%%%%%%%%%%%%%%%%%%%%%%%%%%%%%%%%%%%%%%%%%%%%%%%%%%%%%%%%%%%%%%%%%%%%%%%
%%%%%%%%%%%%%%%%%%%%%%%%%%%%%%%%%%%%%%%%%%%%%%%%%%%%%%%%%%%%%%%%%%%%%%%%%%%%%%%%%%
%%%%%%%%%%%%%%%%%%%%%%%%%%%%%%%%%%%%%%%%%%%%%%%%%%%%%%%%%%%%%%%%%%%%%%%%%%%%%%%%%%
\section{Results}
\label{Results} 
%%%%%%%%%%%%%%%%%%%%%%%%%%%%%%%%%%%%%%%%%%%%%%%%%%%%%%%%%%%%%%%%%%%%%%%%%%%%%%%%%%
%%%%%%%%%%%%%%%%%%%%%%%%%%%%%%%%%%%%%%%%%%%%%%%%%%%%%%%%%%%%%%%%%%%%%%%%%%%%%%%%%%
%%%%%%%%%%%%%%%%%%%%%%%%%%%%%%%%%%%%%%%%%%%%%%%%%%%%%%%%%%%%%%%%%%%%%%%%%%%%%%%%%%
%%%%%%%%%%%%%%%%%%%%%%%%%%%%%%%%%%%%%%%%%%%%%%%%%%%%%%%%%%%%%%%%%%%%%%%%%%%%%%%%%%

Now we are able to calculate the cross sections for elastic and inelastic $J/\psi $ scattering by pseudoscalar and vector mesons using the framework of unitarized coupled channel amplitudes obtained in previous section. In particular, the channels considered are the $J/\psi X$, with $X$ being the mesons associated to the fields introduced in the $P$ and $V$ matrices in Eq.~(\ref{PV}), i.e. the $\pi,K,\eta,\rho,K^\ast,\omega$ mesons. In this context, we use an enlarged coupled channel basis by taking into account the quantum numbers $I^G(J^{PC})$, charm ($C$) and strangeness ($S$) of each channel. Thus, remembering that in present work our interest is only on $s$-wave processes, in Table~\ref{Tab:Canais} it is displayed the channel content in each sector, determined by analyzing the meson pairs with same quantum numbers and with possible transitions among them. Accordingly, the decomposition of these channels involving light and heavy mesons allows us to obtain the coefficients $\xi _{i j}, \chi _{i j } $ and $\zeta _{i j}$ given in Eqs.~(\ref{Eq:CasoVPVP})-(\ref{Eq:CasoVVVV}); they are given in~\ref{Appendix} in an isospin basis.

%\begin{widetext}
\begin{table}[htp]
	\caption{Channel content in each sector. It is shown only relevant channels for S-wave processes.}
\begin{center}
			\begin{tabular}{c|c}
		\hline
		$\mathbf{I^G(J^{PC})}$& $\mathbf{C=S=0}$ \\ 
		\hline
				\hline

		$0^+(0^{++}), 0^-(1^{+-})$ 		& \multirow{2}{*}{$J/\psi J/\psi, \omega J/\psi, \omega \omega, \rho \rho, D_s^\ast \bar D_s^\ast$}\\
		$0^+(2^{++})$ 					& \\ 
		\hline
		%$0^+(0^{++})$ 					& $\pi\pi, \eta\eta, \eta_c\eta, \eta_c \eta_c, D_s \bar D_s$ \\ 
		\multirow{2}{*}{$0^-(1^{+-})$}	& $\pi \rho, \eta \omega, \eta J/\psi, \eta_c \omega, K \bar K^\ast - c.c.,$\\
		& $\eta_c J/\psi, D \bar D^\ast - c.c., D_s \bar D_s^\ast + c.c.$ \\ 
		\hline
		\multirow{2}{*}{$1^-(0^{++})$}	& $\rho \omega, K^\ast \bar K^\ast, \eta \pi, \bar K K$ \\
		& $\rho J/\psi, D^\ast \bar D^\ast, \eta_c \pi, \bar D D$ \\
		\hline
		$1^+(1^{+-}), 1^-(2^{++})$		& $\rho J/\psi, \rho \omega, K^\ast \bar K^\ast, D^\ast \bar D^\ast$\\ 
		\hline
		\multirow{2}{*}{$1^+(1^{+-})$}	& $\pi \omega, \eta \rho, K \bar K^\ast + c.c.$\\
		& $\pi J/\psi, \eta_c \rho, D \bar D^\ast + c.c.$\\
		\hline 	
		\hline
		$\mathbf{I^G(J^{PC})}$	& $\mathbf{C=0,S=1}$\\ 
		\hline
		\hline
      	\multirow{2}{*}{$\frac{1}{2}(0^+)$}		& $K \eta, K \pi, K^* \omega, K^* \rho$ \\
		& $K \eta_c, D_s \bar D, K^* J/\psi, D_s^\ast \bar D^\ast$ \\
		\hline
		\multirow{2}{*}{$\frac{1}{2}(1^+),\frac{1}{2}(2^+)$} 			& $K^* J/\psi, K^* \omega,$ \\
		& $ K^* \rho, D_s^\ast \bar D^\ast$ \\ 
		\hline
		%$1/2(0^+)$ 					& $\pi K, \eta K, \eta_c K, D_s \bar D$ \\ 
		\multirow{2}{*}{$\frac{1}{2}(1^+)$}		& $\pi K^\ast, \eta K^\ast, K \rho, K \omega$\\
		& $\eta_c K^\ast, J/\psi K,\bar D D_s^\ast, \bar D^\ast D_s$\\	\hline
	\end{tabular}	
\end{center}
	\label{Tab:Canais}
\end{table}
%\end{widetext}

We have employed in the computations of the present work the following values for the masses:  $m_{\pi} = 138 $ MeV, $m_{\rho} = 771 $ MeV, $m_K = 495 $ MeV, $m_{\eta} = 548$ MeV, $m_{\omega} = 782 $ MeV, 
$m_{K^{\ast}} = 892 $ MeV, $m_D = 1865 $ MeV, 
$m_{D^{\ast}} = 2008$ MeV, $m_{D_s} = 1968 $ MeV, $m_{D_s ^{\ast}} = 2008$ 
MeV, $m_{\eta_c} = 2979$ MeV,  $m_{J/\psi} = 3097 $ MeV, $m_{L} = 800 $ MeV, $m_{H} = 2050 $ MeV and $m_{H}^\prime = 3000 $ MeV; for the decay constants: $f_{\pi} = 93$ MeV and  $f_{D} = 165$ MeV. We have fixed the free parameters in the loop function, Eq.~(\ref{G2}), as in Ref.~\cite{Gamermann:2007fi}:  setting the scale $\mu  $ to 1.5 GeV, the subtraction constant is adjusted to data taking $a_H (\mu) = -1.55 $ for channels involving at least one heavy meson, and $a_L (\mu) = - 0.8 $ for channels involving only light mesons.

\begin{figure}
	\centering
	\includegraphics[width=.95\linewidth]{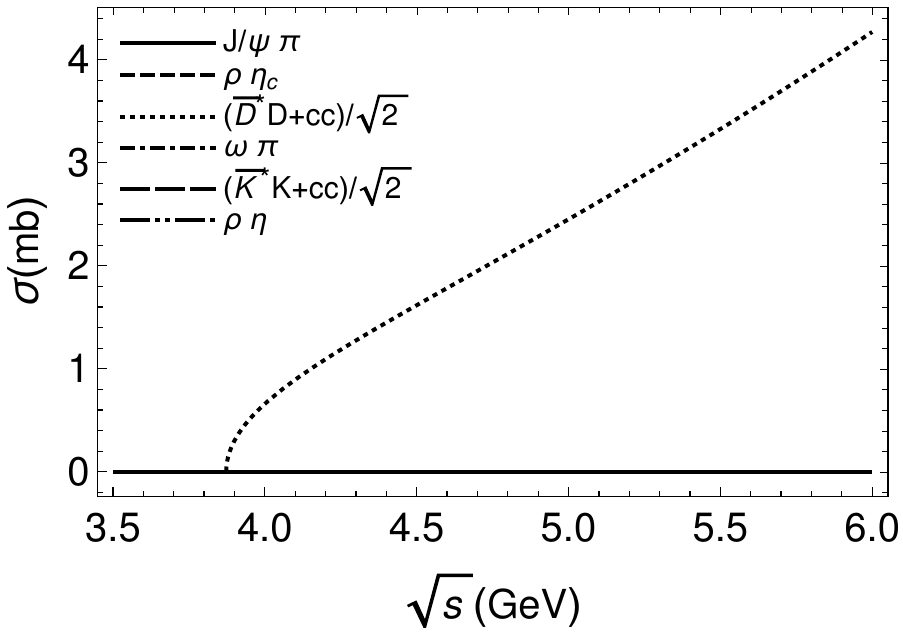} \\ 
	\includegraphics[width=1.0\linewidth]{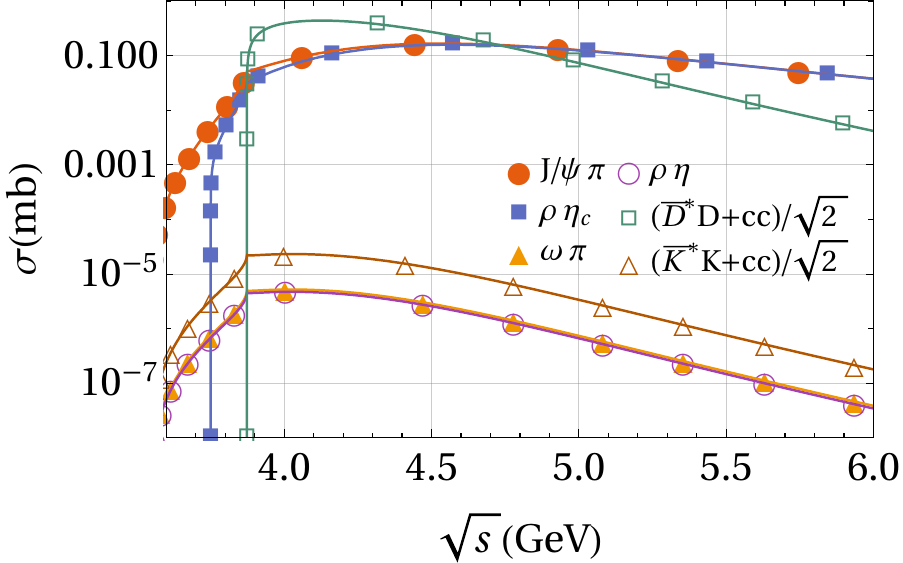}
	\caption{Cross sections for $J/\psi \pi$ scattering into allowed final states as a function of the CM energy $\sqrt{s}$. Top panel: use of tree-level amplitudes.   Bottom panel: use of unitary amplitudes.}
	\label{fig:RelevanceJpi}
\end{figure}

In what follows we present and discuss the cross sections for the $J/\psi$-meson interactions regarding the channel content in each sector, as reproduced Table~\ref{Tab:Canais}. We start by showing in  Fig.~\ref{fig:RelevanceJpi} the most investigated scattering in literature: the cross sections for $J/\psi \pi$ scattering into allowed final states. Particularly, beyond the reactions  $J/\psi \pi \rightarrow  J/\psi \pi, \rho \eta_c, (D \bar D^\ast + c.c.)$, which are also present in Ref.~\cite{PhysRevC.96.045201}, we examine $J/\psi \pi \rightarrow \omega \pi, \rho \eta, ( \bar K^\ast K+c.c. )$ as well. 
Some remarks are worthy of mention when compare them. 
  First, we must take care of the validity of the present treatment: it is valid at low-energy range, since it is employed the lowest order Lagrangian filtered out projecting it onto $s$-wave. Keeping this in mind, we see that 
at three level only the reaction with final state $( D \bar D^\ast + c.c.)$ has non-zero cross section. 
Once the amplitude is unitarized, the meson loops engender non-vanishing cross sections for all reactions, with an universal behavior: they have a peak shortly after the respective threshold, and decrease rapidly or slowly as energy increases, depending on the reaction. In addition, it can be observed that the most relevant processes are those whose final state carries charmed quarks. 
The contributions with final states $J/\psi \pi, \rho \eta_c, ( \bar D^\ast D+c.c.)$ can be regarded as approximately with the same order of magnitude in the energy range under consideration. On the other hand, they are greater than cross sections for $J/\psi \pi \rightarrow \omega \pi, \rho \eta,( \bar K^\ast K+c.c.)$ by about a factor $10^5$, which justifies the neglect of these last reactions for practical purposes. 

Another point we would like to observe is on the comparison of our results with existing literature. In general, the cross section we have obtained for $J/\psi \pi \rightarrow (D \bar D^\ast + c.c.)$ reaction has a comparable or smaller magnitude  at low CM energies than other ones~\cite{Wong:1999zb,Wong:2001td,PhysRevC.58.2994,PhysRevC.61.031902,Braun-Munzinger2000,PhysRevC.62.034903,PhysRevC.63.065201,PhysRevC.63.034901,PhysRevC.68.014903,Oh:2002vg,PhysRevC.68.035208,Maiani:2004py,Maiani:2004qj,DURAES200397,PhysRevC.70.055203,PhysRevC.72.024902,PhysRevD.72.034002,Capella2008,CASSING20011,doi:10.1142/S0218301308010507,PhysRevC.85.064904,MITRA201675,Liu2016,PhysRevC.96.045201,PhysRevC.97.044902}. As CM energy grows, the high-energy behavior of the presented findings show a more pronounced decrease of magnitude of  cross section than those with any kind of control of high-energy behavior. Possible discrepancies can be attributed to the different energy dependence of the adopted formalism describing the interactions; 
 contributions of higher partial waves;  
distinct approach employed to control the high-energy behavior, as in the cases of form-factors; and differing values of coupling constants, masses, cutoffs, ...  Notwithstanding, it is worthy noticing that a faster decreasing for higher CM energies  qualitatively similar to our findings in Ref.~\cite{Oh:2002vg}, which makes use of covariant form-factors.

\begin{figure}
	\centering
	\includegraphics[width=1.0\linewidth]{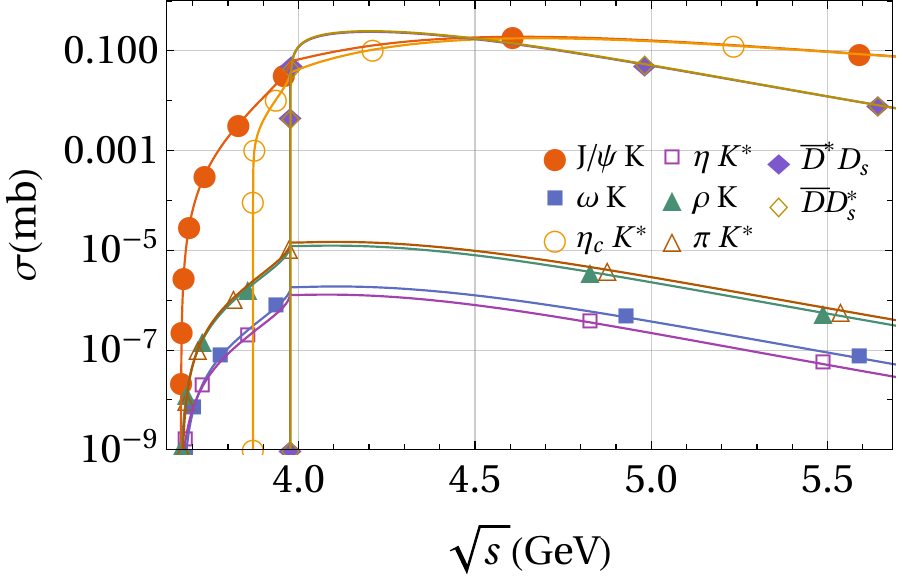} \\ 
	\includegraphics[width=1.0\linewidth]{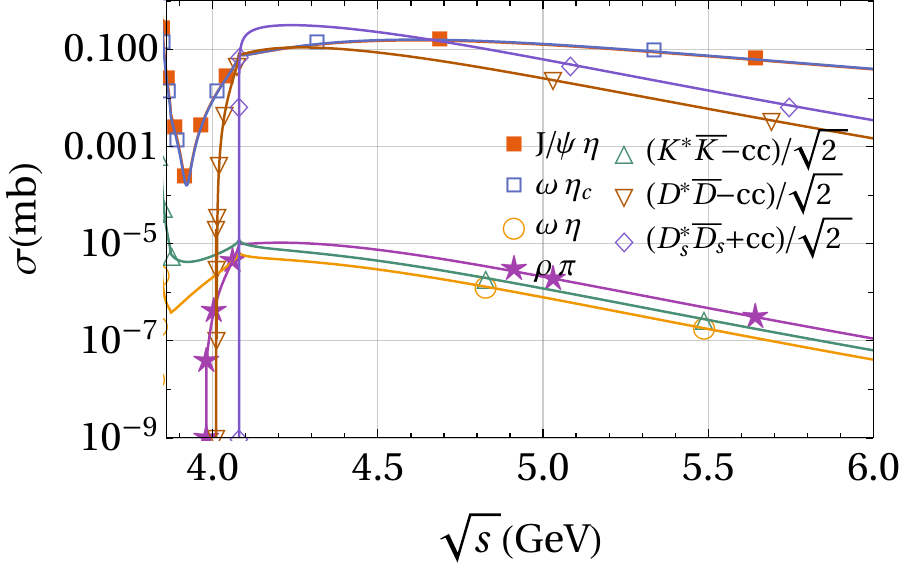}
	\caption{Unitarized cross sections for $J/\psi K $ (top panel) and $J/\psi \eta $ (bottom panel) scatterings into allowed final states as a function of the CM energy $\sqrt{s}$. }
	\label{fig:RelevanceJKeta}
\end{figure}

\begin{figure}
	\centering
	\includegraphics[width=0.92\linewidth]{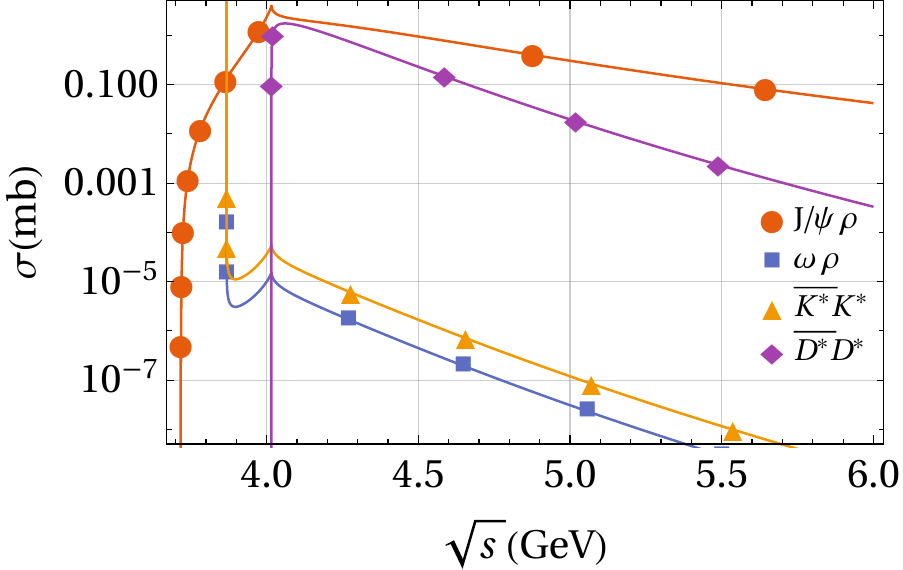} \\ 
	\includegraphics[width=0.92\linewidth]{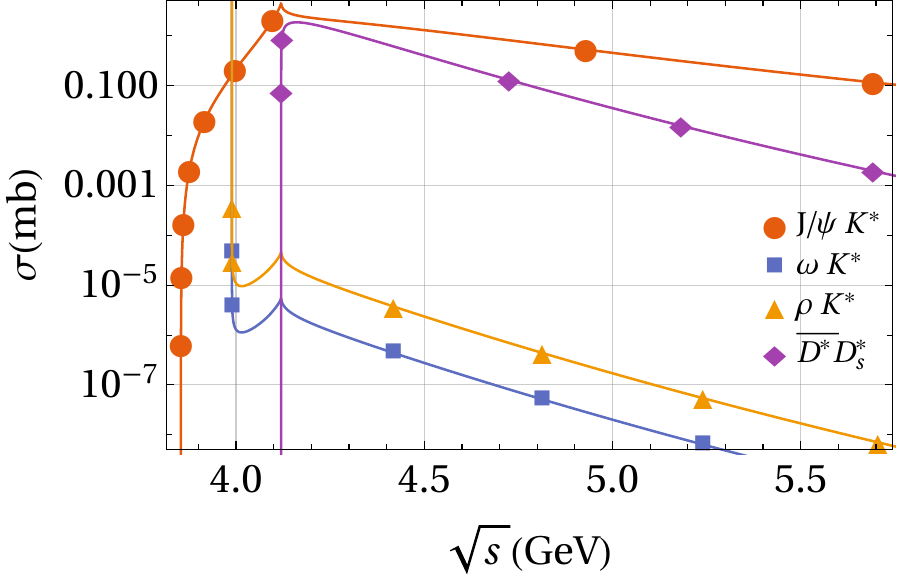} \\
		\includegraphics[width=0.92\linewidth]{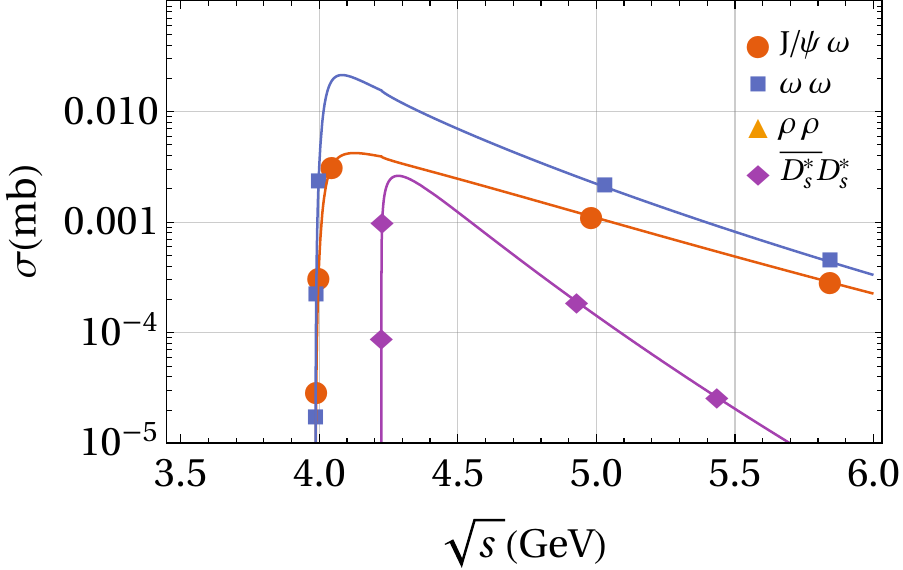}
	\caption{Unitarized cross sections for $J/\psi \rho $ (top panel), $J/\psi K^{\ast} $ (center panel), and $J/\psi \omega $ (bottom panel) scatterings into allowed final states as a function of the CM energy $\sqrt{s}$. }
	\label{fig:RelevanceJrhoKstaromega}
\end{figure}

For completeness, In Figs.~\ref{fig:RelevanceJKeta} and ~\ref{fig:RelevanceJrhoKstaromega} are also plotted the unitarized cross sections for $J/\psi X$ scatterings into allowed final states, with $X$ being the pseudoscalar and vector mesons $K, \eta,\rho, K^{\ast} , \omega $. 
In view of these results, we remark the points below:
\begin{itemize}
    
\item At tree level, before unitarization procedure, only reactions with open charmed mesons in final states (i.e. $J/\psi X \rightarrow \bar D_{(s)}^{(\ast)} D_{(s)}^{(\ast)} + c.c. $) have non-vanishing cross sections, with an uncontrolled behavior with energy. 

\item The unitarized coupled channel amplitudes via the meson loops generate non-vanishing and controlled cross sections, with a peak shortly after the threshold and a decrease with increasing energy. 

\item  In general, reactions with charmed final state are the most relevant contributions for the cross sections, while the other ones have a very small magnitude and are highly suppressed as energy increases. Precisely, most relevant processes are the elastic ones, $J/\psi X \rightarrow J/\psi X$, as well as the inelastic ones with $\eta_c$ and open charmed mesons in final states (($J/\psi X \rightarrow \eta_c Y$) and $J/\psi X \leftrightarrow \bar D_{(s)}^{(\ast)} D_{(s)}^{(\ast)} + c.c. $). 

\item In the case of $J / \psi \omega$ scattering the final state $\rho \rho$ does not appear in the plot, since it is vanishing. The reason is due to the fact that meson loops do not generate allowed combinations for this channel. 

\item In the plots of the cross sections for $ J / \psi $ scattering by vector mesons, we have considered the sum of the situations with different spin contributions ($ J = 0,1,2 $). However, we have restricted ourselves to the $ VV  \rightarrow VV $ processes, because of the negligible contributions of $ VV \rightarrow PP $ ones (see comment below). In this sense, we have not taken into account these latter channels both in the mesonic loops and in the final states.

\item We have employed the lowest order Lagrangian in chiral expansion, with their contributions projected onto $s$-wave. In this sense, higher partial waves would dominate the cross section at greater CM energies  above threshold, which  would modify the faster decreasing of cross sections.

\end{itemize}

\begin{figure}[h]
	\centering
	\includegraphics[width=1.0\linewidth]{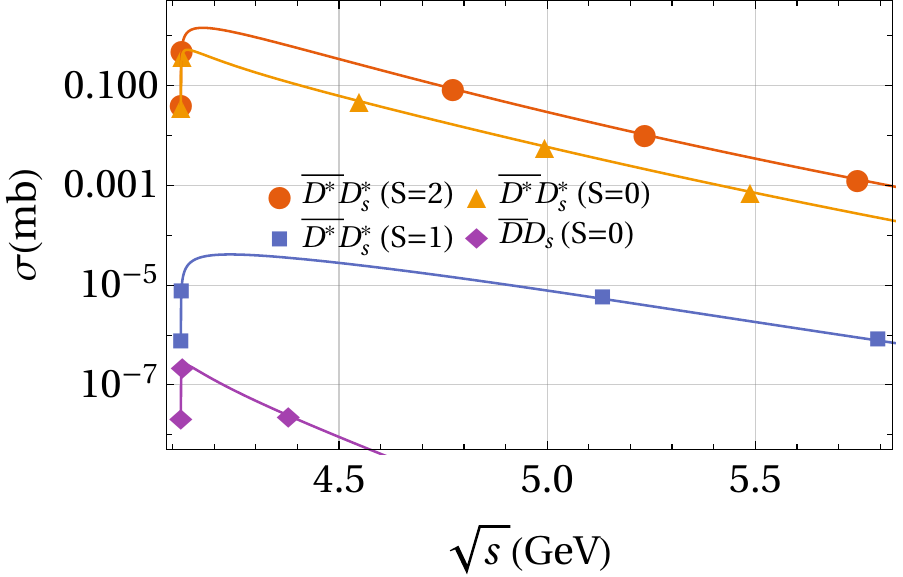}
	\caption{Cross section of $J/\psi$ scattering by the vector meson $K^\ast$ using the unitarized coupled channel approach. It is shown only final states with open charmed mesons. For $\bar D^\ast D^\ast$ there are three combinations of total spin that are exhibited.	}
	\label{fig:DDrelevance1}
\end{figure}

Furthermore, we should add some comments concerning the large suppression of magnitude for the processes $VV \rightarrow PP$. Due to the nature of this interaction, the only way to obtain one reaction of this type is through $s$-channel in Eq.~(\ref{Eq:CasoVVVV}), which is proportional to the term $(t-u)$. In particular, if the $s$-channel is zero, as for the $J/\psi \omega$ scattering, $VV \rightarrow PP$ reactions are forbidden. Nevertheless, the $J/\psi \rho$ and  $J/\psi K^{\ast}$ scatterings have not all $s$-channels being null.
Notwithstanding, notice that at $s$-wave, $(t-u)$ is $(m_1^2-m_2^2)(m_1^{'2}-m_2^{'2})$, where $m_i$ are the masses of the incoming particles and $m_i'$ the masses of the outgoing particles. Consequently, when the incoming or outgoing particles have close masses, the $s$-channel becomes highly suppressed. This effect can be illustrated from the cross sections of the reaction $J/\psi K^\ast$ taking as final states open charmed mesons, as shown in Fig.~\ref{fig:DDrelevance1}. As it can be seen, the contribution of reaction with final state being $(\bar D D_{(s)} + c.c.)$ is largely suppressed with respect to the other ones. This result is relatively reproduced  in the findings of Ref.~\cite{PhysRevC.97.044902}, in which the processes $\sigma(J/\psi K^\ast \rightarrow \bar D ^{\ast} D_{s}^{\ast} + c.c.)$ and $\sigma(J/\psi K^\ast \rightarrow \bar D D_{s} + c.c.)$  have cross sections with amplitudes that differ by about a factor $10^2$.

%\begin{widetext}  
  \begin{center}
	\begin{figure}[h]
		\centering
		\includegraphics[width=1.0\linewidth]{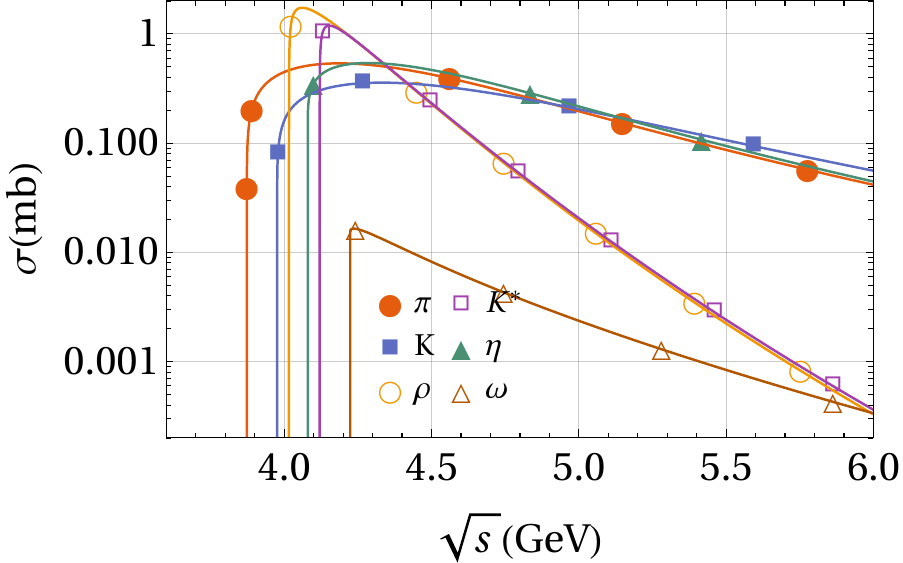}\\
		\caption{Cross-sections as function of center-of-mass energy $\sqrt{s}$ for $J/\psi X $ scattering into all allowed final states; $X$ denotes $\pi,K,\eta,\rho,K^\ast,\omega$ mesons. }
		\label{fig:Relevance_IN}
	\end{figure}
\end{center}
%\end{widetext}
  \begin{center}
	\begin{figure}[h]
		\centering
		\includegraphics[width=1.0\linewidth]{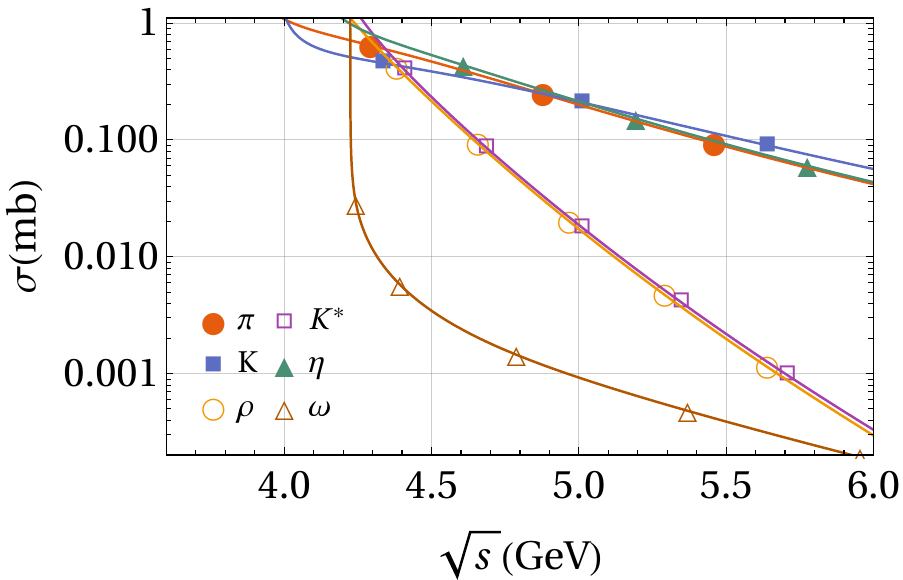}
		\caption{Cross-sections as function of center-of-mass energy $\sqrt{s}$ for inverse reactions discussed in Fig.~\ref{fig:Relevance_IN}.}
		\label{fig:Relevance_OUT}
	\end{figure}
\end{center}

We summarize the results above by estimating the cross sections for the $J/\psi$ with each meson resulting in all possible channels; they are plotted in Fig.~\ref{fig:Relevance_IN}. It is clear that the cross sections involving pseudoscalars $J/\psi P \rightarrow \text{All}$ (where $All$ means the coupled channels to each of the initial state according to Table~\ref{Tab:Canais}) have magnitudes larger than those with vector mesons ($J/\psi V \rightarrow \text{All}$ ). This result is qualitatively in accordance with Ref.~\cite{PhysRevC.97.044902} as well as other works,  always taking care of the validity of the present approach. 

Finally, in Fig.~\ref{fig:Relevance_OUT} is shown the cross sections for inverse reactions discussed in Fig.~\ref{fig:Relevance_IN}, i.e. $ \text{All} \rightarrow J/\psi X$. We notice that the cross sections for direct and inverse processes  can be considered to be approximately of the same order of magnitude: they are between 0.1 and 1 mbarn in the range $4 $ GeV $< \sqrt{s} < 5$ GeV, and are suppressed at high energies.  

Hence, the findings reported above allow us to evaluate the most relevant interactions between the $J/\psi$ resonance and the hadronic medium composed of the lightest mesons, and will be useful for the determination of evolution of $J/\psi$ abundance in high energy collisions, even as for correspondence among other procedures. 

%%%%%%%%%%%%%%%%%%%%%%%%%%%%%%%%%%%%%%%%%%%%%%%%%%%%%%%%%%%%%%%%%%%%%%%%%%%%%%%%%%
%%%%%%%%%%%%%%%%%%%%%%%%%%%%%%%%%%%%%%%%%%%%%%%%%%%%%%%%%%%%%%%%%%%%%%%%%%%%%%%%%%
%%%%%%%%%%%%%%%%%%%%%%%%%%%%%%%%%%%%%%%%%%%%%%%%%%%%%%%%%%%%%%%%%%%%%%%%%%%%%%%%%%
%%%%%%%%%%%%%%%%%%%%%%%%%%%%%%%%%%%%%%%%%%%%%%%%%%%%%%%%%%%%%%%%%%%%%%%%%%%%%%%%%%
\section{Concluding Remarks}
\label{Conclusions} 
%%%%%%%%%%%%%%%%%%%%%%%%%%%%%%%%%%%%%%%%%%%%%%%%%%%%%%%%%%%%%%%%%%%%%%%%%%%%%%%%%%
%%%%%%%%%%%%%%%%%%%%%%%%%%%%%%%%%%%%%%%%%%%%%%%%%%%%%%%%%%%%%%%%%%%%%%%%%%%%%%%%%%
%%%%%%%%%%%%%%%%%%%%%%%%%%%%%%%%%%%%%%%%%%%%%%%%%%%%%%%%%%%%%%%%%%%%%%%%%%%%%%%%%%
%%%%%%%%%%%%%%%%%%%%%%%%%%%%%%%%%%%%%%%%%%%%%%%%%%%%%%%%%%%%%%%%%%%%%%%%%%%%%%%%%%

In this work we have evaluated the interactions of $J/\psi$ with surrounding hadronic medium. We have considered the medium composed of light pseudoscalar mesons ($\pi, K, \eta$) and vector mesons ($\rho, K^\ast, \omega$), and calculated the cross sections for $J/\psi $ scattering by light mesons, as well as their inverse processes. Within the framework of  unitarized coupled channel amplitudes, we have analyzed the magnitude of unitarized cross sections of the different channels, and performed a comparison of our results with existing literature.

The employment of unitarized coupled channel amplitudes via the meson loops have generated non-vanishing and controlled cross sections, including reactions without open charmed mesons in final states which have zero-amplitudes at tree level. Also,  from the results it can be inferred that reactions with charmed final state are the most relevant contributions for the cross sections, while the other ones have a very small magnitude and are highly suppressed as energy increases. Another feature is the negligible contribution of $ VV \rightarrow PP $ processes both in the mesonic loops and in the final states.

Moreover, concerning the estimates of the cross sections for the $J/\psi$ with each meson resulting in all possible channels, they suggest that the scattering $J/\psi P \rightarrow~\text{All}$ have magnitudes larger than those with vector mesons ($J/\psi V \rightarrow \text{All}$ ) in the most range of center-of-mass energy $\sqrt{s}$.

It is relevant to notice the limitations of the present treatment. Since it has been employed the lowest-order Lagrangian in chiral expansion, with their contributions projected onto $s$-wave, therefore in principle the investigation of low-energy range near threshold is valid, despite there are outcomes reported in literature whose higher-energy behavior is qualitatively similar to ours.

Further work is needed to improve these results, in order to perform more precise comparison with predictions made by other phenomenological models. In particular, the analysis of higher partial waves would modify the decreasing of cross sections at greater energies, and will be useful in the determination of evolution of $J/\psi$ abundance in high energy collisions.

%%%%%%%%%%%%%%%%%%%%%%%%%%%%%%%%%%%%%%%%%%%%%%%%%%%%%%%%%%%%%%%%%%%%%%%%%%%%%%%%&&&&
%%%%%%%%%%%%%%%%%%%%%%%%%%%%%%%%%%%%%%%%%%%%%%%%%%%%%%%%%%%%%%%%%%%%%%%%%%%%%%%%%%%%
%%%%%%%%%%%%%%%%%%%%%%%%%%%%%%%%%%%%%%%%%%%%%%%%%%%%%%%%%%%%%%%%%%%%%%%%%%%%%%%%%%%%
%%%%%%%%%%%%%%%%%%%%%%%%%%%%%%%%%%%%%%%%%%%%%%%%%%%%%%%%%%%%%%%%%%%%%%%%%%%%%%%%%%%%
%\begin{acknowledgements}
%
%
%
%We are grateful to F. S. Navarra and M. Nielsen for reading our manuscript and for fruitful discussions. 
%The authors would like to thank the Brazilian funding agencies CNPq (contracts
%
%308088/2017-4 and 400546/2016-7) and FAPESB (contract INT0007/2016) for financial support.
%
%
%\end{acknowledgements} 
%
%%%%%%%%%%%%%%%%%%%%%%%%%%%%%%%%%%%%%%%%%%%%%%%%%%%%%%%%%%%%%%%%%%%%%%%%%%%%%%%%&&&&
%%%%%%%%%%%%%%%%%%%%%%%%%%%%%%%%%%%%%%%%%%%%%%%%%%%%%%%%%%%%%%%%%%%%%%%%%%%%%%%%%%%%
%%%%%%%%%%%%%%%%%%%%%%%%%%%%%%%%%%%%%%%%%%%%%%%%%%%%%%%%%%%%%%%%%%%%%%%%%%%%%%%%&&&&
%%%%%%%%%%%%%%%%%%%%%%%%%%%%%%%%%%%%%%%%%%%%%%%%%%%%%%%%%%%%%%%%%%%%%%%%%%%%%%%%%%%%

\begin{center}
{\bf ACKNOWLEDGMENTS}
\end{center}
%%%%%%%%%%%%%%%%
%%%%%%%%%%%%%%%%

We are grateful to F. S. Navarra and M. Nielsen for reading our manuscript and for fruitful discussions. 
The authors thank CNPq (Brazilian Agencies) for financial support. 
L.M.A. would like to thank the funding agencies CNPq (contracts
308088/2017-4 and 400546/2016-7) and FAPESB (contract INT0007/2016) for financial support.

%\bibliography{library}{}
%
%\bibliographystyle{apsrev4-1}
%
%\appendix

%%%%%%%%%%%%%%%%%%%%%%%%%%%%%%%%%%%%%%%%%%%%%%%%%%%%%%%%%%%%%%%%%%%%%%%%%%%%%%%%&&&&
%%%%%%%%%%%%%%%%%%%%%%%%%%%%%%%%%%%%%%%%%%%%%%%%%%%%%%%%%%%%%%%%%%%%%%%%%%%%%%%%%%%%
%%%%%%%%%%%%%%%%%%%%%%%%%%%%%%%%%%%%%%%%%%%%%%%%%%%%%%%%%%%%%%%%%%%%%%%%%%%%%%%%&&&&
%%%%%%%%%%%%%%%%%%%%%%%%%%%%%%%%%%%%%%%%%%%%%%%%%%%%%%%%%%%%%%%%%%%%%%%%%%%%%%%%%%%%
\appendix
%%%%%%%%%%%%%%%%%%%%%%%%%%%%%%%%%%%%%%%%%%%%%%%%%%%%%%%%%%%%%%%%%%%%%%%%%%%%%%%%&&&&
%%%%%%%%%%%%%%%%%%%%%%%%%%%%%%%%%%%%%%%%%%%%%%%%%%%%%%%%%%%%%%%%%%%%%%%%%%%%%%%%%%%%
%%%%%%%%%%%%%%%%%%%%%%%%%%%%%%%%%%%%%%%%%%%%%%%%%%%%%%%%%%%%%%%%%%%%%%%%%%%%%%%%&&&&
%%%%%%%%%%%%%%%%%%%%%%%%%%%%%%%%%%%%%%%%%%%%%%%%%%%%%%%%%%%%%%%%%%%%%%%%%%%%%%%%%%%%
\section[Appendix]{$\xi _{i j}, \chi _{i j } $ and $\zeta _{i j}$ coefficients}

\label{Appendix}

%\subsection{$\xi _{i j}, \chi _{i j } $ and $\zeta _{i j}$ coefficients \label{Coefficients}}

The decomposition of the channels involving light and heavy mesons allows us to obtain the coefficients $\xi _{i j}, \chi _{i j } $ and $\zeta _{i j}$ given in Eqs.~(\ref{Eq:CasoVPVP})-(\ref{Eq:CasoVVVV}). Here we summarize the values that they must assume with the choice of a proper isospin basis, and according to the type of mesons involved in relevant channels ($VP \rightarrow VP, VV \rightarrow PP$, and $VV \rightarrow VV$). Here, we denote 
\begin{eqnarray}
\gamma & = & \left(\frac{m_L}{m_H} \right)^2, \nonumber \\
\psi & = & - \frac{1}{3} + \frac{4}{3}\left(\frac{m_L}{m_H ^{\prime}} \right)^2 , %\nonumber \\
%\kappa & = & \left(\frac{f_{\pi}}{f_D}\right),  
\label{sb} 
\end{eqnarray}
where the values of these quantities are given in Section~\ref{Results}. 

\subsection{$VP \rightarrow VP \; (S=1)$ }

The non-vanishing $VP \rightarrow VP$ scatterings in Eq.~(\ref{Eq:CasoVPVP}) are only $s$-wave processes. The coefficients $\xi _{i j}$ are shown in the tables below. 
%\begin{widetext}
 \begin{center}
\begin{tabular}{c||c|c|c|c|c|c|c|c}
	\hline
	\multicolumn{9}{c}{$\mathbf{C=S=0}, \mathbf{I^G(J^{PC})} = 0^-(1^{+-})$}\\
	\hline
		\hline
\multirow{2}{*}{}	Channel &$J/\psi\eta_c $&$J/\psi\eta $&$\omega\eta_c $&$\omega\eta $&$\rho\pi $ &   $\bar K^\ast K  $ & $ \bar D^\ast D$&$\bar D_s^\ast D_s $\\
 &  &  &  &  &   & 	$ - c.c. $ & $ - c.c. $&$ + c.c.$\\
	\hline
		\hline
	$J/\psi\eta_c $& $0 $&$ 0 $&$ 0 $&$ 0 $&$ 0 $&$ 0 $&$ \frac{4 \gamma   }{3} $&$ \frac{\sqrt{8} \gamma  }{3} $  \\
	 \hline
	$J/\psi\eta $& $0 $&$ 0 $&$ 0 $&$ 0 $&$ 0 $&$ 0 $&$ \frac{\sqrt{2} \gamma  }{3}   $&$ \frac{-2 \gamma  }{3}$ \\
	 \hline
	$\omega\eta_c $& $0 $&$ 0 $&$ 0 $&$ 0 $&$ 0 $&$ 0 $&$ \frac{\sqrt{2} \gamma  }{3}   $&$ \frac{-2 \gamma  }{3} $\\
	 \hline
	$\omega\eta $& $0 $&$ 0 $&$ 0 $&$ 0 $&$ 0 $&$ \frac{-3}{2} $&$ \frac{\gamma   }{6} $&$ \frac{\sqrt{2} \gamma  }{3}   $\\
	 \hline
	$\rho\pi $& $0 $&$ 0 $&$ 0 $&$ 0 $&$ 2 $&$ \frac{\sqrt{3}}{2} $&$ \frac{- \sqrt{3} \gamma   }{2} $&$ 0 $\\
	 \hline
	$\bar K^\ast K - c.c.$& $0 $&$ 0 $&$ 0 $&$ \frac{-3}{2} $&$ \frac{\sqrt{3}}{2} $&$ \frac{3}{2} $&$ \frac{-\gamma  }{2}$&$ \frac{-\gamma   }{\sqrt{2}} $\\
	 \hline
	$ \bar D^\ast D - c.c.$& $\frac{4 \gamma   }{3} $&$ \frac{\sqrt{2}\gamma  }{3}    $&$ \frac{\sqrt{2}\gamma   }{3}   $&$ \frac{\gamma   }{6} $&$ \frac{-\sqrt{3}\gamma  }{2}    $&$ \frac{-\gamma  }{2} $&$ \frac{  (\psi +2)}{2}  $&$ \frac{1 }{\sqrt{2}}$ \\
	 \hline
	$\bar D_s^\ast D_s + c.c. $& $\frac{\sqrt{8}\gamma  }{3}   $&$ \frac{-2 \gamma  }{3} $&$ \frac{-2 \gamma  }{3}$&$ \frac{\sqrt{2} \gamma   }{3}  $&$ 0 $&$ \frac{-\gamma   }{\sqrt{2}} $&$ \frac{1 }{\sqrt{2}} $&$ \frac{  (\psi +1)}{2} $ \\
	 \hline	
\end{tabular}
\end{center}

  \begin{center}
\begin{tabular}{c||c|c|c|c|c|c}
	\hline
	\multicolumn{7}{c}{$\mathbf{C=S=0}, \mathbf{I^G(J^{PC})} = 1^+(1^{+-}) $}\\
	\hline
	\hline
	Channels &$J/\psi \pi$&$\omega \pi$&$\rho \eta_c$&$\rho \eta$&$ \bar K^\ast  K + c.c.$&$ \bar D^\ast D+ c.c.$\\
	\hline						     
	\hline
	$J/\psi \pi$& 0 & 0 & 0 & 0 & 0 & $-\sqrt{\frac{2}{3}} \gamma  $  \\
	\hline
	$\omega \pi$& 0 & 0 & 0 & 0 & $-\frac{\sqrt{3}}{2}$ & $-\frac{\gamma   }{2 \sqrt{3}} $\\
	\hline
	$\rho \eta_c$& 0 & 0 & 0 & 0 & 0 &$ -\sqrt{\frac{2}{3}} \gamma  $  \\
	\hline
	$\rho \eta$& 0 & 0 & 0 & 0 & $-\frac{\sqrt{3}}{2}$ &$ -\frac{\gamma   }{2 \sqrt{3}}$ \\
	\hline
	$ \bar K^\ast  K+c.c.$& 0 &$ -\frac{\sqrt{3}}{2}$ & 0 &$ -\frac{\sqrt{3}}{2}$ & $\frac{1}{2} $& $\frac{\gamma   }{2} $\\
	\hline
	$ \bar D^\ast D+c.c.$& $-\sqrt{\frac{2}{3}} \gamma   $ &$ -\frac{\gamma   }{2 \sqrt{3}}$ &$ -\sqrt{\frac{2}{3}} \gamma  $  & $-\frac{\gamma   }{2 \sqrt{3}}$ & 
	$\frac{\gamma   }{2}$ &$ \frac{  \psi }{2} $\\
	\hline
\end{tabular}
\end{center}

\begin{center}
\begin{tabular}{c||c|c|c|c|c|c|c|c}
	\hline
	\multicolumn{9}{c}{$\mathbf{C=0, S=1}, \mathbf{I^G(J^{PC})} = 1/2(1^+)$}\\
	\hline
	\hline
	Channels &$ J/\psi K$&$\omega K $&$K^\ast \eta_c $&$K^\ast \eta $&$\rho K $&$ K^\ast \pi$&$ \bar D^\ast D_s$ &$\bar D D_s^\ast$\\
	\hline
	\hline
	$ J/\psi K$& $0 $&$ 0 $&$ 0 $&$ 0 $&$ 0 $&$ 0 $&$ \frac{\gamma   }{\sqrt{3}} $&$ \frac{\gamma   }{\sqrt{3}} $\\
	 \hline 
	$\omega K $& $0 $&$ 0 $&$ 0 $&$ -\frac{3}{4} $&$ 0 $&$ \frac{3}{4} $&$ \frac{\gamma   }{2 \sqrt{6}} $&$ \frac{-\gamma   }{\sqrt{6}}$ \\
	\hline
	$K^\ast \eta_c $& $0 $&$ 0 $&$ 0 $&$ 0 $&$ 0 $&$ 0 $&$ \frac{\gamma   }{\sqrt{3}} $&$ \frac{\gamma   }{\sqrt{3}}$ \\
	 \hline
	$K^\ast \eta $ &$0 $&$ \frac{-3}{4} $&$ 0 $&$ 0 $&$ \frac{3}{4} $&$ 0 $&$ \frac{-\gamma   }{\sqrt{6}} $&$ \frac{\gamma   }{2 \sqrt{6}} $\\
	 \hline
	$\rho K $& $0 $&$ 0 $&$ 0 $&$ \frac{3}{4} $&$ 1 $&$ \frac{1}{4} $&$ \frac{-\sqrt{3}\gamma  }{\sqrt{8}} $&$ 0$ \\
	 \hline
	$ K^\ast \pi$& $0 $&$ \frac{3}{4} $&$ 0 $&$ 0 $&$ \frac{1}{4} $&$ 1 $&$ 0 $&$\frac{-\sqrt{3}\gamma  }{\sqrt{8}} $ \\
	 \hline
	$ \bar D^\ast D_s$ & $\frac{\gamma   }{\sqrt{3}} $&$ \frac{\gamma   }{2 \sqrt{6}} $&$ \frac{\gamma   }{\sqrt{3}} $&$ \frac{-\gamma   }{\sqrt{6}} $&$\frac{-\sqrt{3}\gamma  }{\sqrt{8}} $&$ 0 $&$ \frac{  \psi }{2} $&$ 0$ \\
	 \hline
	$\bar D D_s^\ast$ & $\frac{\gamma   }{\sqrt{3}} $&$ \frac{-\gamma   }{\sqrt{6}} $&$ \frac{\gamma   }{\sqrt{3}} $&$ \frac{\gamma   }{2 \sqrt{6}} $&$ 0 $&$\frac{-\sqrt{3}\gamma  }{\sqrt{8}}  $&$ 0 $&$ \frac{  \psi }{2} $\\
	\hline
\end{tabular}
\end{center}

%\end{widetext}

\subsection{$ VV \rightarrow VV $  and $ VV \rightarrow PP$}

The non-vanishing $VP \rightarrow VP$ scatterings in Eq.~(\ref{Eq:CasoVPVP}) are only $s$-wave processes. The coefficients $\xi _{i j}$ are shown in the tables below. 

As it is shown in Eq.~(\ref{Eq:CasoVVVV}), the $VV \rightarrow VV$ reactions can occur via ($s,t,u$)-processes. Therefore, we exhibit in the tables below the coefficients $(\zeta _{i j} ^{(s)},\zeta _{i j} ^{(t)},\zeta _{i j} ^{(u)})$. 

For processes $VV \rightarrow PP$ the coefficients are obtained just by replacing the vector pair in final state of $VV \rightarrow VV$ by the respective pseudoscalar pair $PP$ in $SU(4)$ basis (i.e. $J/\psi K^*$ by $\eta_c K $, and so on). Notice, however, that $VV \rightarrow PP$ scatterings are proportional to $t-u$, see Eq.~\eqref{Eq:CasoVVPP}. Hence, the coefficients $\chi _{i j} $ are equal to $\zeta _{i j} ^{(s)}$, i.e. they are the first coefficients in the tables below. 

%\begin{widetext}
\begin{center}
\begin{tabular}{c||c|c|c|c}
	\hline
	\multicolumn{5}{c}{$\mathbf{C=S=0}, \mathbf{I^G(J^{PC})} = 1^+(1^{+-}), 1^-(2^{++}); I_z = +1 $}\\
	\hline
	\hline
	Channel &$J/\psi \rho$&$ \omega \rho$& $\bar K^\ast  K^\ast$&$ \bar D^\ast D^\ast$\\
	\hline
	\hline
	$J/\psi \rho$& $0,0,0 $&$ 0,0,0 $&$ 0,0,0 $&$ 0,\frac{-\gamma   }{\sqrt{3}},\frac{-\gamma   }{\sqrt{3}} $\\
		\hline
	$ \omega \rho$& $0,0,0 $&$ 0,0,0 $&$ 0,\frac{-\sqrt{3}}{\sqrt{8}},\frac{-\sqrt{3}}{\sqrt{8}} $&$ 0,\frac{-\gamma   }{2 \sqrt{6}},\frac{-\gamma   }{2 \sqrt{6}}$ \\
	\hline
	$\bar K^\ast  K^\ast$& $0,0,0 $&$ 0,\frac{-\sqrt{3}}{\sqrt{8}},\frac{-\sqrt{3}}{\sqrt{8}} $&$ \frac{1}{2},\frac{1}{2},0 $&$ \frac{-1 }{4},\frac{\gamma   }{8},\frac{3 \gamma   }{8} $\\
	\hline
	$ \bar D^\ast D^\ast$& $0,\frac{-\gamma   }{\sqrt{3}},\frac{-\gamma   }{\sqrt{3}} $&$ 0,\frac{-\gamma   }{2 \sqrt{6}},\frac{-\gamma   }{2 \sqrt{6}} $&$ \frac{-1 }{4},\frac{\gamma   }{8},\frac{3 \gamma   }{8} $&$ \frac{1 }{4},\frac{  (2 \psi +1)}{8},\frac{  (2 \psi -1)}{8}  $\\
	\hline
\end{tabular}
\end{center}

\begin{center}
	\begin{tabular}{c||c|c|c|c}
		\hline
		\multicolumn{5}{c}{$\mathbf{C=S=0}, \mathbf{I^G(J^{PC})} = 1^+(1^{+-}), 1^-(2^{++}); I_z = 0 $}\\
		\hline
	\hline
		Channel &$J/\psi \rho$&$ \omega \rho$& $\bar K^\ast  K^\ast$&$ \bar D^\ast D^\ast$\\
	\hline
	\hline
		$J/\psi \rho$& $0,0,0 $&$ 0,0,0 $&$ 0,0,0 $&$ 0,\frac{-\gamma   }{\sqrt{3}},\frac{-\gamma   }{\sqrt{3}} $\\
	\hline
		$ \omega \rho$& $0,0,0 $&$ 0,0,0 $&$ 0,\frac{-\sqrt{3}}{\sqrt{8}},\frac{-\sqrt{3}}{\sqrt{8}} $&$ 0,\frac{-\gamma   }{2 \sqrt{6}},\frac{-\gamma   }{2 \sqrt{6}}$ \\
	\hline
		$\bar K^\ast  K^\ast$& $0,0,0 $&$ 0,\frac{-\sqrt{3}}{\sqrt{8}},\frac{-\sqrt{3}}{\sqrt{8}} $&$ \frac{1}{4},\frac{1}{4},0 $&$ 0,\frac{\gamma   }{4},\frac{\gamma   }{4} $\\
	\hline
		$ \bar D^\ast D^\ast$& $0,\frac{-\gamma   }{\sqrt{3}},\frac{-\gamma   }{\sqrt{3}} $&$ 0,\frac{-\gamma   }{2 \sqrt{6}},\frac{-\gamma   }{2 \sqrt{6}} $&$0,\frac{\gamma   }{4},\frac{\gamma   }{4} $&$ 0,\frac{ \psi}{4},\frac{ \psi}{4}  $\\
		\hline
	\end{tabular}
\end{center}

\begin{center}
	\begin{tabular}{c||c|c|c|c}
		\hline
		\multicolumn{5}{c}{$\mathbf{C=S=0}, \mathbf{I^G(J^{PC})} = 1^+(1^{+-}), 1^-(2^{++}); I_z = -1 $}\\
		\hline
	\hline
		Channel &$J/\psi \rho$&$ \omega \rho$& $\bar K^\ast  K^\ast$&$ \bar D^\ast D^\ast$\\
	\hline
	\hline
		$J/\psi \rho$& $0,0,0 $&$ 0,0,0 $&$ 0,0,0 $&$ 0,\frac{-\gamma   }{\sqrt{3}},\frac{-\gamma   }{\sqrt{3}} $\\
	\hline
		$ \omega \rho$& $0,0,0 $&$ 0,0,0 $&$ 0,\frac{-\sqrt{3}}{\sqrt{8}},\frac{-\sqrt{3}}{\sqrt{8}} $&$ 0,\frac{-\gamma   }{2 \sqrt{6}},\frac{-\gamma   }{2 \sqrt{6}}$ \\
	\hline
		$\bar K^\ast  K^\ast$& $0,0,0 $&$ 0,\frac{-\sqrt{3}}{\sqrt{8}},\frac{-\sqrt{3}}{\sqrt{8}} $&$ \frac{1}{2},\frac{1}{2},0 $&$ \frac{-1 }{2},0,\frac{\gamma  }{2} $\\
	\hline
		$ \bar D^\ast D^\ast$& $0,\frac{-\gamma  }{\sqrt{3}},\frac{-\gamma   }{\sqrt{3}} $&$ 0,\frac{-\gamma   }{2 \sqrt{6}},\frac{-\gamma   }{2 \sqrt{6}} $&$ \frac{- 1}{2},0,\frac{\gamma   }{2} $&$ \frac{ 1}{2},\frac{  \psi}{2},0  $\\
		\hline
	\end{tabular}
\end{center}

\begin{center}
\begin{tabular}{c||c|c|c|c|c}
	\hline
	\multicolumn{6}{c}{$\mathbf{C=S=0}, \mathbf{I^G(J^{PC})} = 0^+(0^{++}), 0^-(1^{+-}),0^+(2^{++}) $}\\
	\hline
	\hline
	Channels &$J/\psi J/\psi$&$ J/\psi \omega$&$\omega \omega$&$\rho \rho$&$\bar D_s^\ast D_s^\ast $\\
	\hline
	\hline
	$J/\psi J/\psi$&$ 0,0,0 $&$ 0,0,0  $&$ 0,0,0  $&$ 0,0,0  $&$ 0,\frac{2 \gamma   }{3},\frac{2 \gamma   }{3}$ \\
	\hline
	$ J/\psi \omega$&$ 0,0,0  $&$ 0,0,0  $&$ 0,0,0  $&$ 0,0,0  $&$ 0,-\frac{\sqrt{2} \gamma  }{3},-\frac{\sqrt{2} \gamma  }{3}  $ \\
	\hline
	$\omega \omega$&$ 0,0,0  $&$ 0,0,0  $&$ 0,0,0  $&$ 0,0,0  $&$ 0,\frac{\gamma   }{3},\frac{\gamma   }{3}$ \\
	\hline
	$\rho \rho$&$ 0,0,0  $&$ 0,0,0  $&$ 0,0,0  $&$ 0,2,2 $&$ 0,0,0  $\\
	\hline
	$\bar D_s^\ast D_s^\ast $&$ 0,\frac{2 \gamma   }{3},\frac{2 \gamma   }{3} $&$ 0,-\frac{\sqrt{2} \gamma  }{3},-\frac{\sqrt{2} \gamma  }{3}   $&$0, \frac{\gamma   }{3},\frac{\gamma   }{3} $&$ 0,0,0 $&$ 0,\frac{  (\psi +1)}{4},\frac{  (\psi +1)}{4} $\\
	\hline
\end{tabular}
\end{center}

\begin{center}
\begin{tabular}{c||c|c|c|c}
	\hline
	\multicolumn{5}{c}{$\mathbf{C=0, S=1}, \mathbf{I^G(J^{PC})} = 1/2(0^+),1/2(1^+),1/2(2^+); I_z=+1/2 $}\\
	\hline
	\hline
	Channels &$J/\psi K^*$&$\omega K^*$&$\rho K^*$&$\bar D^\ast D_s^\ast$\\
		\hline
	\hline
	$J/\psi K^*$	&$0,0,0 $&$ 0,0,0 $&$ 0,0,0 $&$ 0,\frac{\gamma   }{\sqrt{3}},\frac{\gamma   }{\sqrt{3}} $\\
		\hline
	$\omega K^*$	&$0,0,0 $&$ \frac{3}{4},0,\frac{-3}{4} $&$ \frac{-1}{4},0,\frac{1}{4} $&$ \sqrt{\frac{3}{8}}  ,\frac{\gamma   }{2 \sqrt{6}},\frac{-\gamma   }{\sqrt{6}} $\\
		\hline
	$\rho K^*$	&$0,0,0 $&$ \frac{-1}{4},0,\frac{1}{4} $&$ \frac{1}{12},-\frac{1}{3},-\frac{5}{12} $&$ \frac{-1 }{2 \sqrt{6}},\frac{-\gamma   }{2 \sqrt{6}},0 $\\
		\hline
	$\bar D^\ast D_s^\ast$	&$0,\frac{\gamma   }{\sqrt{3}},\frac{\gamma   }{\sqrt{3}} $&$ \sqrt{\frac{3}{8}}  ,\frac{\gamma   }{2 \sqrt{6}},\frac{-\gamma   }{\sqrt{6}} $&$ \frac{-1 }{2 \sqrt{6}},\frac{-\gamma   }{2 \sqrt{6}},0 $&$ \frac{1 }{2},\frac{  \psi }{2},0 $\\
		\hline
\end{tabular}
\end{center}

\vspace{5pt}

\begin{center}
\begin{tabular}{c||c|c|c|c}
	\hline
	\multicolumn{5}{c}{$\mathbf{C=0, S=1}, \mathbf{I^G(J^{PC})} = 1/2(0^+),1/2(1^+),1/2(2^+); I_z=-1/2 $}\\
	\hline
	\hline
	Channels &$J/\psi K^*$&$\omega K^*$&$\rho K^*$&$\bar D^\ast D_s^\ast$\\
	\hline
	\hline
	$J/\psi K^*$	&$0,0,0 $&$ 0,0,0 $&$ 0,0,0 $&$ 0,\frac{\gamma   }{\sqrt{3}},\frac{\gamma   }{\sqrt{3}} $\\
	\hline
	$\omega K^*$	&$0,0,0 $&$ \frac{3}{4},0,\frac{-3}{4} $&$ \frac{3}{4},0,\frac{-3}{4} $&$ \sqrt{\frac{3}{8}}  ,\frac{\gamma   }{2 \sqrt{6}},\frac{-\gamma   }{\sqrt{6}} $\\
	\hline
	$\rho K^*$	&$0,0,0 $&$ \frac{3}{4},0,\frac{-3}{4} $&$ \frac{3}{4},1,\frac{1}{4} $&$ \frac{\sqrt{3} }{\sqrt{8}},\frac{\sqrt{3}\gamma   }{\sqrt{8}},0 $\\
	\hline
	$\bar D^\ast D_s^\ast$	&$0,\frac{\gamma   }{\sqrt{3}},\frac{\gamma   }{\sqrt{3}} $&$ \sqrt{\frac{3}{8}}  ,\frac{\gamma   }{2 \sqrt{6}},\frac{-\gamma   }{\sqrt{6}} $&$ \frac{\sqrt{3} }{\sqrt{8}},\frac{\sqrt{3}\gamma   }{\sqrt{8}},0 $&$ \frac{1 }{2},\frac{  \psi }{2},0 $\\
	\hline
\end{tabular}
\end{center}
%\end{widetext}

%\section{More figures}
%
%
%
%For the sake of readability we have not shown all figures during the main text. However, for completeness, we expose then here.
%
%
%
%\subsection{Consequence of unitarization}
%
%
%
%\begin{figure}
%
%\centering
%
%\includegraphics[width=\linewidth]{NewFigs/SUP02}
%
%\includegraphics[width=\linewidth]{NewFigs/SUP03}
%
%\caption{Suppression $J/\psi X \rightarrow \text{All}$ for both the non-unitarized amplitudes (gray line) and unitarized amplitudes (black lines).}
%
%\label{fig:SUP02}
%
%\end{figure}
%
%\FloatBarrier
%
%\subsection{..}
%

%%%%%%%%%%%%%%%%%%%%%%%%%%%%%%%%%%%%%%%%%%%%%%%%%%%%%%%%%%%%%%%%%%%%%%%%%%%%%%%%&&&&
%%%%%%%%%%%%%%%%%%%%%%%%%%%%%%%%%%%%%%%%%%%%%%%%%%%%%%%%%%%%%%%%%%%%%%%%%%%%%%%%%%%%
%%%%%%%%%%%%%%%%%%%%%%%%%%%%%%%%%%%%%%%%%%%%%%%%%%%%%%%%%%%%%%%%%%%%%%%%%%%%%%%%&&&&
%%%%%%%%%%%%%%%%%%%%%%%%%%%%%%%%%%%%%%%%%%%%%%%%%%%%%%%%%%%%%%%%%%%%%%%%%%%%%%%%%%%%

%\bibliography{library}{}
%
%\bibliographystyle{article}
%

%

%%%%%%%%%%%%%%%%%%%%%%%%%%%%%%%%%%%%%%%%%%%%%%%%%%%%%%%%%%%%%%%%%%%%%%%%%%%%%%%%&&&&
%%%%%%%%%%%%%%%%%%%%%%%%%%%%%%%%%%%%%%%%%%%%%%%%%%%%%%%%%%%%%%%%%%%%%%%%%%%%%%%%%%%%
%%%%%%%%%%%%%%%%%%%%%%%%%%%%%%%%%%%%%%%%%%%%%%%%%%%%%%%%%%%%%%%%%%%%%%%%%%%%%%%%&&&&
%%%%%%%%%%%%%%%%%%%%%%%%%%%%%%%%%%%%%%%%%%%%%%%%%%%%%%%%%%%%%%%%%%%%%%%%%%%%%%%%%%%%

\end{document}